# Inverse Design of Metamaterials with Manufacturing-Guiding Spectrum-to-Structure Conditional Diffusion Model

Jiawen Li,[1,6] Jiang Guo,[1,2,6,*] Yuanzhe Li,[3] Zetian Mao,[1] Jiaxing Shen,[2] Tashi Xu,[2] Diptesh Das,[1] Jinming He,[4] Run Hu,[4] Yaerim Lee,[2] Koji Tsuda,[1,5,**] and Junichiro Shiomi[2,3,5,7,***]

[1]Department of Computational Biology and Medical Science, The University of Tokyo, Japan
[2]Department of Mechanical Engineering, The University of Tokyo, Japan
[3]Institute of Engineering Innovation, The University of Tokyo, Japan
[4]School of Energy and Power Engineering, Huazhong University of Science and Technology, China
[5]RIKEN Center for Advanced Intelligence Project, Japan
[6]These authors contributed equally
[7]Lead contact
*Correspondence: guo.jiang@photon.t.u-tokyo.ac.jp
**Correspondence: tsuda@k.u-tokyo.ac.jp
***Correspondence: shiomi@photon.t.u-tokyo.ac.jp

## SUMMARY

Metamaterials are artificially engineered structures that manipulate electromagnetic waves, having optical properties absent in natural materials. Recently, machine learning for the inverse design of metamaterials has drawn attention. However, the highly nonlinear relationship between the metamaterial structures and optical behaviour, coupled with fabrication difficulties, poses challenges for using machine learning to design and manufacture complex metamaterials. Herein, we propose a general framework that implements customised spectrum-to-shape and size parameters to address one-to-many metamaterial inverse design problems using conditional diffusion models. Our method exhibits superior spectral prediction accuracy, generates a diverse range of patterns compared to other typical generative models, and offers valuable prior knowledge for manufacturing through the subsequent analysis of the diverse generated results, thereby facilitating the experimental fabrication of metamaterial designs. We demonstrate the efficacy of the proposed method by successfully designing and fabricating a free-form metamaterial with a tailored selective emission spectrum for thermal camouflage applications.

## KEYWORDS

metamaterials, inverse design, conditional generation, diffusion model, thermal camouflage

## INTRODUCTION

In recent years, metamaterials have found potential applications in various advanced technologies because their unique engineered structures can efficiently manipulate electromagnetic waves, tailoring their optical, thermal radiation, or terahertz properties [1–4]. Metasurface materials typically consist of periodically arranged sub-wavelength structural unit cells, allowing for the customization of the desired optical response by designing unit cells with freely shaped two-dimensional geometric patterns and size parameters [5]. Although designing unit cell structures based on physical intuition[6] has proven effective in devices with simple functionality, the development of large-scale complex metamaterials has remained limited. However, inverse design has emerged as a promising strategy to address the increasing complexity of device functionality and degrees of freedom in design [7]. Optimization algorithms and deep neural networks are commonly employed in inverse design, achieving notable successes in metamaterial design [8–12]. However, as the design freedom of metamaterials increases, the highly nonlinear electromagnetic wave equations between the structure and optical response lead to nonunique solutions, making it challenging to map a given response to multiple structures accurately, potentially resulting in suboptimal designs [13–15]. Additionally, the high-dimensional search spaces increase computational demands and can lead to local minima, whereas large data points can introduce instability in neural network training, further complicating the design process [16].

The emergence of generative models has provided researchers with opportunities to explore more complex metamaterials, particularly those with geometric patterns and higher degrees of freedom. Popular generative models, such as the variational autoencoder (VAE) [17] and generative adversarial network (GAN) [18], have been widely applied in the inverse design of metamaterials. For example, Ma et al. [9] employed a conditional VAE to generate metamaterial patterns under predefined spectrum response conditions. The encoder maps the metamaterial patterns and their optical responses from a high-dimensional space to a low-dimensional latent space, and the decoder maps the latent space and optical responses back to the original patterns to achieve conditional generation. Similarly, Liu et al. [19] utilized evolutionary strategies to optimize the latent space of metamaterial patterns in a traditional VAE for inverse design purposes. The inverse design of metamaterials based on GAN involves adversarial training between a generator and a discriminator to generate structures with predefined optical responses [20,21]. The optical responses and random noise are inputs into the generator to produce patterns. The discriminator learns the differences between the generated patterns and the original ones. Although VAE and GAN have been extensively studied in metamaterial inverse design, they present limitations that lead to manufacturing challenges. VAE often produces low-resolution patterns with discontinuous pixels because of the compression of intricate features into a lower-dimensional latent space, the risk of posterior collapse [22], and approximation errors [23]. Further, the lack of structure details impacts the design accuracy of metamaterial structures, and the discontinuity of generation potentially reduces manufacturing practicality. GAN, which can generate higher-quality images, is unstable and prone to mode collapse during training, often resulting in uniform and less diverse patterns [24]. This leads to suboptimal designs that fail to exploit the potential of metamaterials fully, requiring extensive tuning and multiple training runs, thereby increasing computational costs and prolonging the design process.

Recently, diffusion models [25] have gained attention in computer vision fields because of their ease of training and stability, enabling the generation of high-fidelity and diverse images [26–29]. Based on their success in image generation tasks, their applications have expanded to the inverse design of materials. For instance, Zhang et al. [30] developed a conditional diffusion model with a global classifier-free guidance scale for metasurface inverse design, successfully generating symmetrical H-shaped metasurface patterns that met specific spectrum and size parameters. However, this enforced symmetric patterns and the combined use of spectrum and size parameters together as conditions restrict the actual design space, and the size parameters are not inversely designed in their approach, making the inverse design impractical. Additionally, their model with noise-dependent guidance scales can lead to spatial inconsistencies [31], particularly under complex spectrum conditions. Different regions in metamaterial patterns exhibit varying spectrum intensities, and the global guidance scale can cause pixel-level discrepancies in the conditional generation, making it difficult for the model to capture the spatial relationship between patterns and spectra. To address the data requirements in data-driven models, Kim et al. [32] employed an unconditional diffusion model to generate patterns that mimic the distribution of the training data, aiming to augment the dataset for training a model to predict the spectrum from the patterns. Although this approach can expand the dataset, it risks overfitting because of the similar distributions between the augmented and existing data. The reliance on pseudo-labelling in the augmented data can introduce inaccuracies in model predictions. Moreover, the semi-supervised learning method lacks learning in the nonlinear relationships between spectrum and structures.

Based on the successful application of the diffusion model in image generation in the computer vision field, we propose a robust inverse design framework for metamaterials named DiffMeta to perform inverse design of metamaterials. DiffMeta comprises a fixed forward diffusion process and a trainable reverse diffusion process, as shown in Figure 1. The top-layer structures are treated as 2D geometric patterns. In the forward diffusion process, Gaussian noise is incrementally introduced to the geometric patterns of the metamaterial, transforming them into noisy patterns. The reverse process aims to generate data with a distribution like that of the training samples, conditioned on the associated spectra. This process begins with a normal distribution and uses a neural network model trained to remove noise iteratively from the desired spectrum, guided by a spectrum-conditioning mechanism based on cross-attention. This enables

DiffMeta to capture the complex relationship between the patterns and spectra, learning the nonlinear relationship between structures and spectra. This effectively generates diverse metamaterial structures tailored to specific spectrum requirements. The learned joint features of the patterns and spectral properties can be used to predict corresponding size parameters. Additionally, we incorporated practical manufacturing constraints into our dataset construction to ensure that the designed structures are feasible for experimental fabrication, such as feature size, edge smoothing, layer thickness, curvature radius, shape continuity and so on. Our results demonstrate that DiffMeta can generate high-quality and diverse metamaterial structures with desired optical properties, outperforming existing techniques on multiple evaluation metrics. The accuracy and diversity of DiffMeta's designs not only facilitate conditional generation from spectrum to structures but also provide size error tolerance ranges and shape robustness distributions. This effectively guides the practical manufacturing process by identifying critical design parameters and tailoring fabrication strategies based on detailed structural insights.

To illustrate the versatility and potential of our framework, we present a case study wherein we designed and fabricated a free-form metamaterial structure with tailored emission behavior for thermal camouflage applications. Thermal camouflage conceals the thermal signatures of objects by tuning either temperature, emissivity, or both to match the flux of thermal radiance with the surrounding thermal radiation background, focusing on minimizing thermal radiation within the atmospheric transparent window of 3–5 and 8–13 µm, which can be readily detected by infrared (IR) detectors. Current options for thermal camouflage design include multilayer thin films [33–38], metal-insulator-metal (MIM) structures [39–44] with a standard symmetric metal shape as the top metasurface, and multi-resonance designs that combine different sizes and shapes of metasurfaces based on MIM structures [45,46]. Thin-film designs struggle to suppress high-order resonances or peaks in the short wavelength range, which is unfavorable for thermal camouflage. MIM designs typically exhibit narrowband thermal emissions within the nontransparent atmospheric range, making them suboptimal for thermal heat dissipation. In contrast, multi-resonance designs, which combine multiple resonance elements to broaden the thermal emission peak, can mitigate the narrow bandwidth problem. However, they still exhibit low-emissivity dips when different peaks overlap, and the fabricated structures are prone to degradation with increasing surrounding temperatures. Here, we utilize both surface plasmons and magnetic polaritons by incorporating a complex irregular metal shape on top of a typical MIM structure to achieve broadband thermal emission properties in the nontransparent thermal dissipation band, while suppressing thermal emission elsewhere. The complex irregular shape is crucial for supporting more surface plasmon modes by resonating in different patterns and further coupling with each other or with magnetic resonances. The DiffMeta-designed metamaterial structure exhibits nearly 80% thermal emission heat flux of a blackbody at 180 °C, and the emission apart from the targeted band is largely suppressed, with emissivity nearly 0.1 in the 3–5 and 8–15 µm ranges, making it ideal for thermal camouflage applications.

## *Inverse Design of Metamaterials*
## DiffMeta

DiffMeta is based on a conditional diffusion model proposed by Rombach et al. [28]. Unconditional diffusion models[25] generate images by learning from a set of training images. In the forward process, an input image is increasingly contaminated by noise over multiple steps. During training, a neural network is trained to estimate the added noise vector from the contaminated images. In the reverse process, the neural network produces a less contaminated image from a more contaminated one. By repeatedly denoising a white Gaussian noise image, a novel clean image is obtained.

In addition to the size parameters, our task was to learn a mapping from the spectrum to the pattern. In the forward process of our conditional diffusion model, DiffMeta, the noise applied at each step is modulated by the spectrum via a U-Net[47]-like neural network in which cross-attention layers are incorporated to integrate the spectrum and pattern information (Figure 1B). In the reverse process, a contaminated image and the target spectrum are fed to the U-Net to create a less contaminated image (Figure 1C). Unlike the unconditional model, the noise and denoising processes are influenced by the spectrum through an encoder. DiffMeta receives a white noise image and the target spectrum, producing a pattern corresponding to the target spectrum (Figure 1C). To generate the size parameters alongside the spectrum, we added a convolutional neural network [48] (CNN) that maps an intermediate layer of the U-Net to the parameters. This additional network was trained jointly with the other parts. See Supplemental Information Section S3 for methodological details. We then conducted a post-design analysis to gain prior knowledge for manufacturing by analysing a variety of diverse generated metamaterial structures, aiming to improve efficiency and reduce costs.

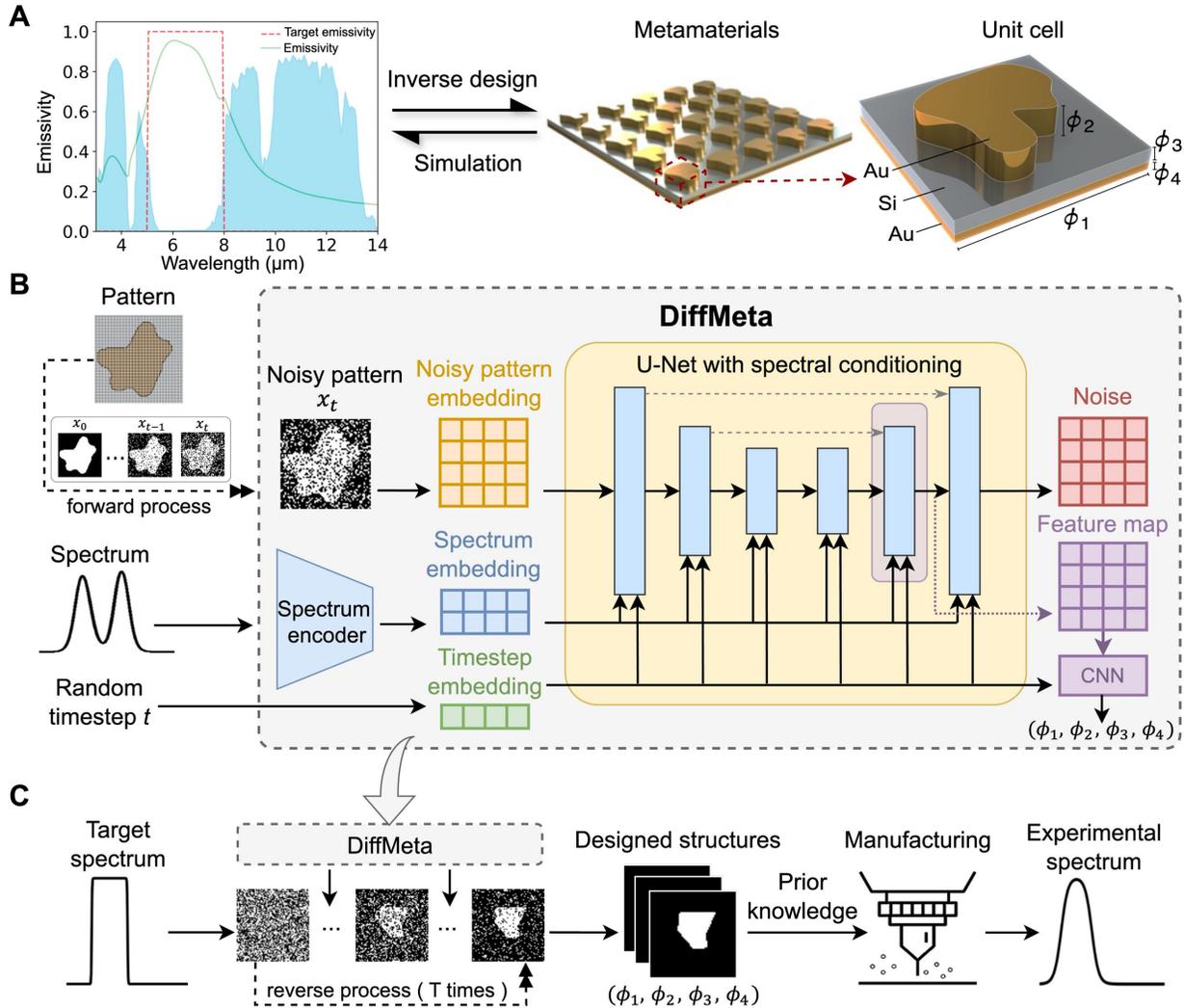

**Figure 1. Metamaterial inverse design workflow by conditional diffusion model.**

(A) Metamaterial design. The red line in the spectrum represents the target spectrum. The MIM tri-layer metamaterial structure used in this study consists of a free-form gold top layer, an amorphous silicon insulator thin film middle layer, and a uniform gold thin film bottom layer. The size parameters ($\phi$) include pitch size ($\phi_1$), top pattern height ($\phi_2$), middle dielectric spacer height ($\phi_3$), and bottom reflector height ($\phi_4$). (B) DiffMeta training phase. The grey block depicts the main structure of the DiffMeta model. U-Net with a spectral conditioning mechanism learns the noise added to patterns in the forward process under the associated spectrum conditions. The spectra are mapped to intermediate layers of the U-Net by a pre-trained spectral encoder. An ensemble of CNN is trained to predict size parameters using pixel-level representations from the middle layer of the U-Net decoder (purple block on U-Net).

(C) DiffMeta generation phase. The trained DiffMeta predicts noise at each timestep under a target spectrum, and the reverse process removes the noise from the previous state over T times. Finally, DiffMeta generates new metamaterial structures that can be manufactured with prior knowledge from the analysis of diverse generated metamaterial structures to obtain spectra that approximate the target spectrum.

### Performance of DiffMeta

Both the conditional VAE and conditional GAN have been employed in the inverse design of metamaterials in recent years. Conditional VAE and conditional GAN models provided by Ma et al. [49] predict patterns from the spectrum, but they do not predict the size parameters. To enable a fair comparison with DiffMeta, the conditional GAN and VAE were modified to allow the prediction of size parameters (see Supplemental Information Section S4.3 for details). These models are referred to hereafter as GANMeta and VAEMeta, respectively. We conducted a grid search of the hyperparameters for each model to select the optimal configuration, ensuring a fair comparison (see Supplemental Information Section S4.4 for details).

**Table 1. Inverse design performance comparison for DiffMeta, VAEMeta, and GANMeta.**

| Model | Pattern Error | Size Parameter Error | Spectrum Error |
|---|---|---|---|
| DiffMeta | **0.0047** | **0.0136** | **0.0619** |
| VAEMeta | 0.0232 | 0.0158 | 0.0650 |
| GANMeta | 0.0159 | 0.0692 | 0.1547 |

In our test set, we have 6000 spectra simulated by RCWA from the ground-truth pattern-parameter pairs. DiffMeta and the other methods were evaluated in terms of inverse design, i.e., how closely a generated pattern-parameter pair resembles the ground truth. DiffMeta, VAEMeta, and GANMeta generated one pattern–parameter pair for a given input spectrum. First, the generated pattern and size parameters were evaluated separately as the pattern error and size parameter errors listed in Table 1. The pattern error and size parameter errors indicate the differences between the generated and ground-truth values for the pattern and size parameters, respectively. Additionally, we compared the input spectrum and the spectrum simulated by RCWA from the generated pair as the spectrum error in Table 1, indicating the difference between the input spectrum and the spectrum simulated from the generated pair. The best results are highlighted in bold. See Supplemental Information Section S2 for these evaluation criteria. In all cases, DiffMeta performed better than VAEMeta and GANMeta. GANMeta showed the worst spectrum error, implying that its inverse design ability was significantly worse than the other two. VAEMeta was close to DiffMeta in spectrum error but not in pattern error, indicating that VAEMeta could not recover reasonable shapes, as discussed in the following sections.

### Pattern Generation Quality

Figure 2 shows examples of the generated patterns. The first row shows the original ground-truth patterns, followed by the three rows depicting the patterns generated by DiffMeta, VAEMeta, and GANMeta, respectively. It is observed that DiffMeta reconstructs high-quality patterns regardless of complexity. In contrast, VAEMeta tends to produce rough outlines of the patterns, possibly constrained by the prior distribution during reconstruction, making it difficult to preserve the details of the original patterns. Particularly in the case of complex patterns, the reconstructed patterns with discontinuous pixels may lack sharp boundaries and exhibit blurriness or distortion, leading to compromised designs. Although GANMeta can generate high-quality patterns for simpler shapes, it faces difficulties in capturing intricate details in complex patterns and sometimes produces discontinuous patterns.

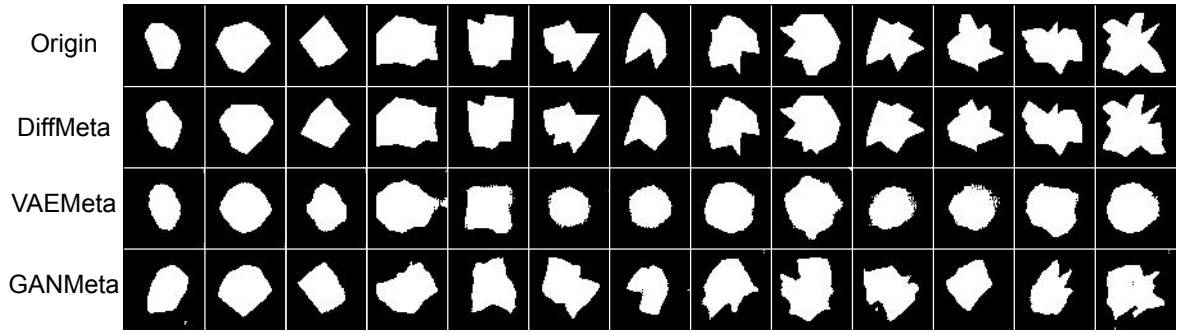

**Figure 2. Examples of generated patterns.**
Top row: Original patterns. Last three rows: Patterns generated by DiffMeta, VAEMeta, and GANMeta, respectively.

## Generated Spectra

Figure 3 shows three input spectra, and the spectra simulated from the generated pattern-parameter pair. The input spectra were deliberately chosen to represent low, medium, and high complexities in the dataset; see the spectrum complexity distribution in Figure S1. According to the spectral complexity proposed by Li et al. [50], the selected spectrum entropies are 0.84, 0.91, and 0.96, respectively. The increasing complexity of the spectrum corresponds to more intricate patterns, posing challenges for generative models in learning the data distribution. In the first example, where the spectrum was relatively simple, both DiffMeta and GANMeta generated patterns that closely approximated the original patterns. As previously analyzed, DiffMeta and VAEMeta demonstrated more accurate size predictions. Looking at the simulated spectra of the generated metamaterials by the three models, all three successfully generated metamaterials that meet the conditions with only minor deviations. In the second example, with a more complex spectrum, DiffMeta generated a simulated spectrum that almost matched the input spectrum. While VAEMeta and GANMeta exhibited some errors, they still demonstrated a conditional generation that approximated the input spectrum.

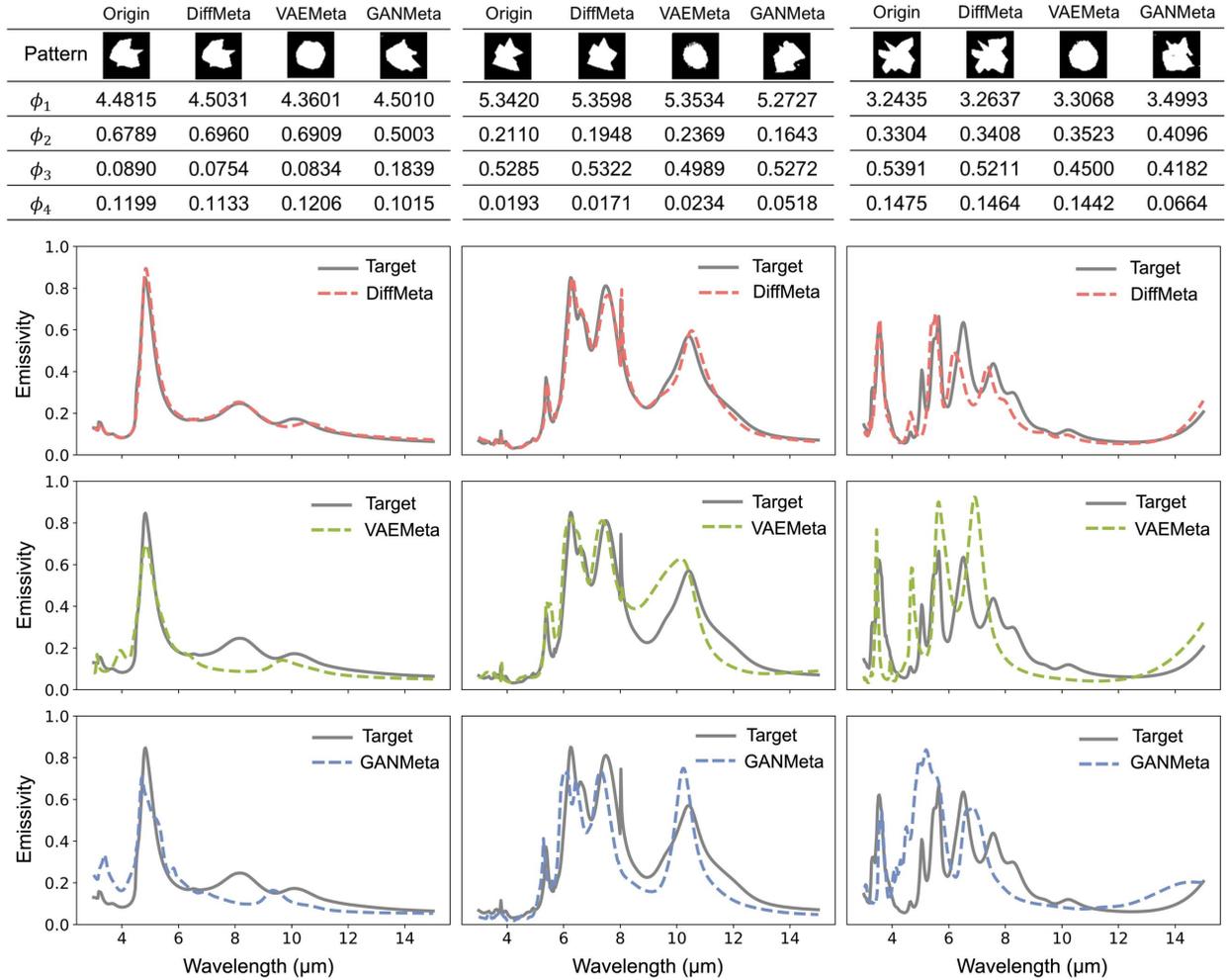

| | Origin | DiffMeta | VAEMeta | GANMeta | | Origin | DiffMeta | VAEMeta | GANMeta | | Origin | DiffMeta | VAEMeta | GANMeta |
|---|---|---|---|---|---|---|---|---|---|---|---|---|---|---|
| Pattern | | | | | | | | | | | | | | |
| $\phi_1$ | 4.4815 | 4.5031 | 4.3601 | 4.5010 | | 5.3420 | 5.3598 | 5.3534 | 5.2727 | | 3.2435 | 3.2637 | 3.3068 | 3.4993 |
| $\phi_2$ | 0.6789 | 0.6960 | 0.6909 | 0.5003 | | 0.2110 | 0.1948 | 0.2369 | 0.1643 | | 0.3304 | 0.3408 | 0.3523 | 0.4096 |
| $\phi_3$ | 0.0890 | 0.0754 | 0.0834 | 0.1839 | | 0.5285 | 0.5322 | 0.4989 | 0.5272 | | 0.5391 | 0.5211 | 0.4500 | 0.4182 |
| $\phi_4$ | 0.1199 | 0.1133 | 0.1206 | 0.1015 | | 0.0193 | 0.0171 | 0.0234 | 0.0518 | | 0.1475 | 0.1464 | 0.1442 | 0.0664 |

**Figure 3. Comparison of input conditional spectrum and simulated spectrum of generated metamaterials by the models.**
The solid gray line represents the input spectrum derived from original patterns and size parameters. Examples display increasing free-form patterns with corresponding increasingly complex spectra.

In the third example, presenting patterns with the highest degrees of freedom and the most complex conditional spectrum, DiffMeta produces a simulated spectrum closely resembling the input spectrum. However, VAEMeta and particularly GANMeta perform less effectively under these challenging conditions. Although GAN generates images that resemble the originals and are of higher quality than those from VAE, it is highly unstable because of the dynamic balance required between the generator and discriminator during adversarial training. Introducing conditional generation further complicates GANMeta training, negatively affecting conditional generation performance. In the inverse design of metamaterials, GANMeta may generate visually appealing patterns, but its instability is a notable drawback for practical metamaterial design. In contrast, DiffMeta proves robust and effective in handling intricate design requirements, faithfully preserving spectral characteristics dictated by input conditions.

## Diversity in Generation

Unlike traditional optimization methods in metamaterial design, generative models like DiffMeta can create diverse solutions from the same input. Generating diverse solutions benefits users by allowing them to choose a desired solution based on practical manufacturing considerations. Thus, we generated samples five times using the three models with different random seeds under a specific input spectrum, as shown in Figure 4. Our observations revealed distinctly different behaviour of the models. GANMeta exhibited significant mode collapse issues, resulting in overly homogeneous and insufficiently diverse samples. VAEMeta, while providing better conditional generation accuracy than GANMeta, still produced blurry and similar patterns as a result of approximation errors, compromising the precision required for structure design. In contrast, DiffMeta effectively generated high-quality and diverse patterns. The results indicated that DiffMeta provides one-to-many design solutions that may be viable and beneficial for practical fabrication.

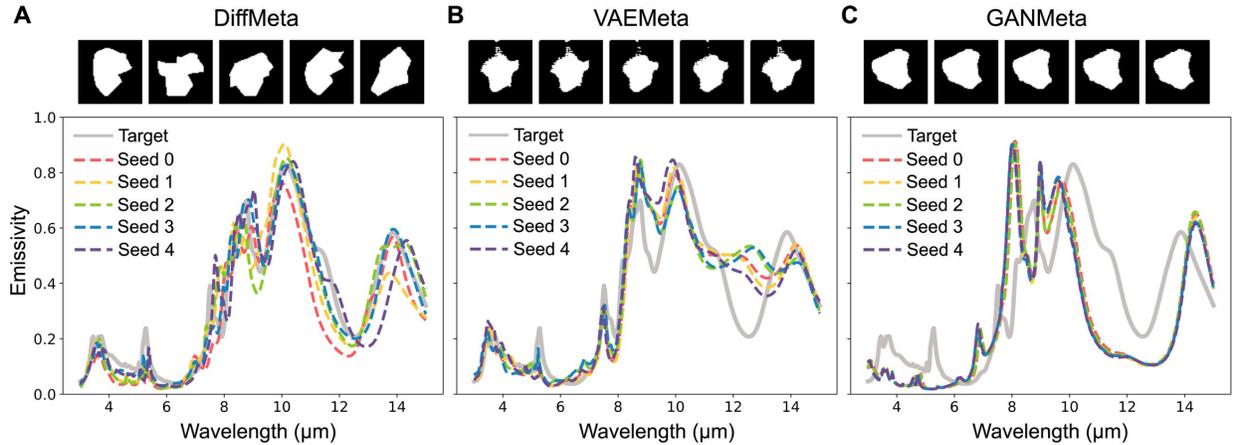

**Figure 4. Illustration of a conditional spectrum yielding multiple solutions.**
Samples were generated five times using three models of (A) DiffMeta, (B) VAEMeta, and (C) GANMeta respectively with different random seeds under a specific input spectrum.

### *Case Study: Thermal Camouflage Metamaterial*

In addition, we designed a case study for thermal camouflage applications, setting the target emissivity property to a step top-hat function, as shown by the red line in the spectrum in Figure 1A. This function aims for nearly 100% emission in the 5–8 μm wavelength range, corresponding to the nontransparent atmospheric window, while maintaining nearly 0% emission in the 3–5 and 8–14 μm ranges, which correspond to the transparent and detectable atmospheric windows.

### Structures Designed with DiffMeta

Depending on the target spectrum, DiffMeta uses different random seeds to generate diverse metamaterial structures. Their corresponding spectra are simulated using a spectrum simulator, as shown in Figure 5A. The results demonstrate that a range of pattern structures and size parameters can achieve a simulated spectrum closely approximating the target spectrum, effectively providing one-to-many solutions for specific design problems. We observed that the predicted values for two key size parameters, ($\phi_1$) and ($\phi_2$), show high concentrations around specific values, suggesting lower variance (Figure 5B). The designed shape distribution is illustrated in Figure 5C, indicating the key shape features for spectrum generation.

Conventional optimization algorithms, such as Bayesian optimization, deep neural networks, or genetic algorithms, are deterministic approaches that typically yield a single optimal solution or a narrow range of solutions[51–53]. However, these solutions may not be practical for real-world fabrication or applications because of hidden constraints or unforeseen conditions [54–56]. In contrast, conditional diffusion models, probabilistic in nature, can generate diverse and varied solutions that meet multiple requirements simultaneously, offering a flexible and cost-effective approach [57–59]. Conditional diffusion models can be valuable for decision making, particularly in developing strategies that efficiently satisfy multiple constraints and support sophisticated planning tasks[60–62].

By analysing 1000 random sampling results from DiffMeta under the target spectrum, the normalized distribution of the design parameters for the target spectrum is obtained. The pitch size ($\phi_1$) and height ($\phi_2$) of the top shape pattern are tightly clustered, implying precise control and consistency in fabrication. In contrast, the middle dielectric spacer height ($\phi_3$) and bottom reflector height ($\phi_4$) show a broader range of values, indicating larger tolerance in the fabrication process.

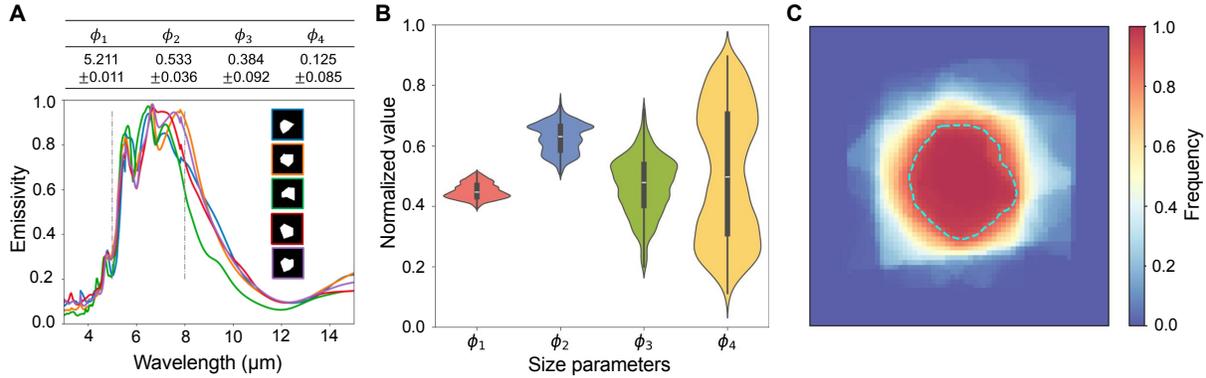

**Figure 5. Analysis of parameters and patterns generated by DiffMeta.**
(A) Simulated spectra of metamaterial structures generated by DiffMeta with different random seeds.
(B) Normalized distribution of design parameters for the target spectrum based on 1000 randomly sampled results.
(C) Distribution heatmap of design patterns for the target spectrum based on 1000 randomly sampled results. The blue dashed line highlights regions where over 90% of patterns are concentrated.

The size parameter analysis is consistent with physical understanding: the pitch size ($\phi_1$) and height ($\phi_2$) affect the surface plasmon resonance peak position and electromagnetic field distribution. The middle dielectric spacer height ($\phi_3$) is responsible for magnetic resonance excitation, requiring a middle region of sufficient thickness to accommodate magnetic resonance excitation, whereas coupling with the top surface plasmon resonance limits its thickness. The bottom reflector height ($\phi_4$) is used to block light transmission, so it has only a minimal thickness requirement, allowing a large variation range. Figure 5C shows the distribution heatmap of the design patterns conditioned on the target spectrum, based on 1000 randomly sampled results. The blue dashed line highlights the regions where over 90% of the generated patterns are concentrated. The heat map reveals that most of the sampled patterns share similar key features. Direct write lithography (DWL) was selected over electron beam lithography (EBL) for metamaterial manufacturing, considering the relatively large size scale and smooth edge distribution of the shape. DWL can effectively handle large size shapes, speed up manufacturing, and reduce costs, making it a potential option for scaling up. By focusing more on those key shape features, the trial and error of the experimental parameters can be reduced, improving manufacturing efficiency.

### Experimental Realization

The generated metamaterial structures were experimentally fabricated for verification. Figure S3 illustrates the step-by-step fabrication process of the metamaterial structure. The morphological characteristics of the fabricated metamaterial structure were examined by scanning electron microscopy (SEM; Hitachi Regulus8230), as shown in Figure 6A. The insets provide magnified views of the precise shape contours. Because of the inherent resolution constraints of the laser writing process, the resulting structures exhibited a degree of rounding effects. For a comparative analysis, the inset illustrates both the original design and the Gaussian-kernel-filtered rounded shape.

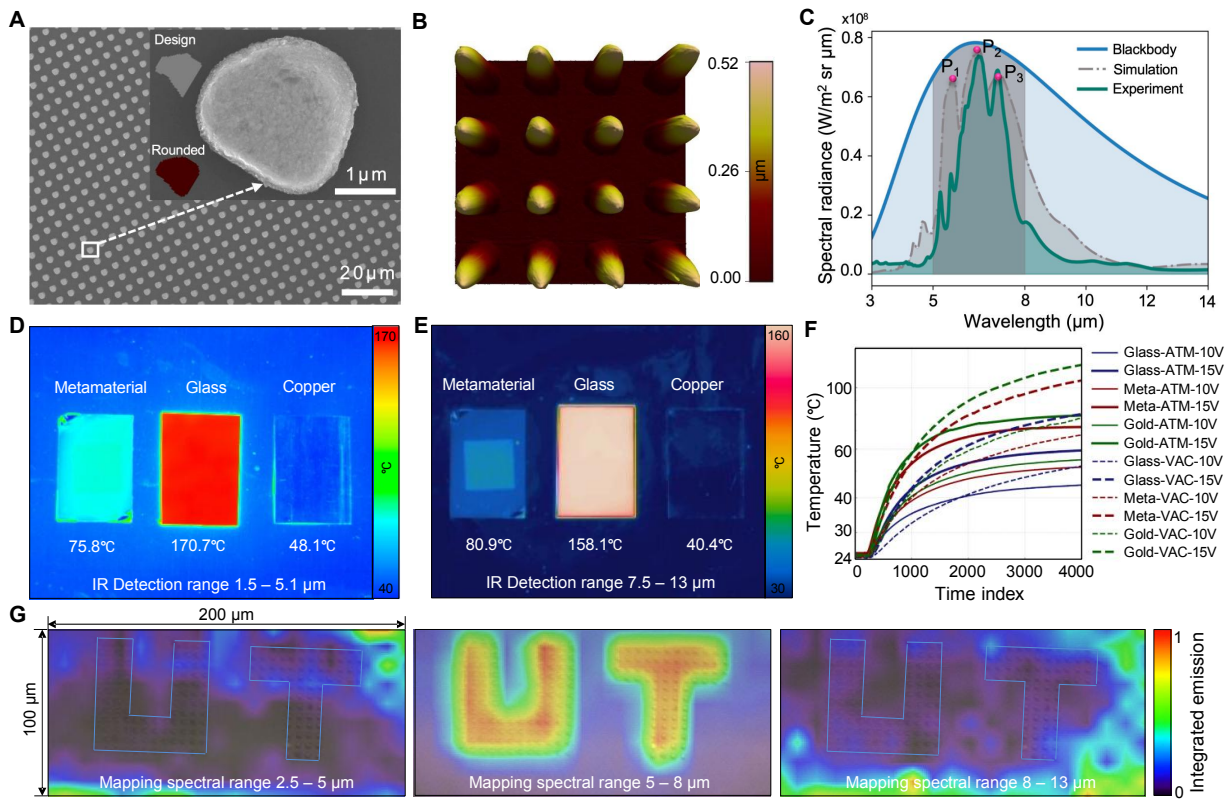

**Figure 6. Metamaterial structure characterization.**

(A) Scanning electron microscope (SEM) image of the metamaterial structure, with the inset showing both the original and rounded designs.

(B) Atomic force microscopy (AFM) image of the top free-form shape metamaterial structure.

(C) Spectral radiance of the blackbody radiation and the simulation and experimental emissivity results of the designed metamaterial structure at 180°C with three major resonant emissivity peaks.

Thermal IR image for (D) Mid-wave IR detector (1.5–5.1 µm) and (E) long-wave IR detector (7.5–13 µm) showing the metamaterial structure, glass, and copper reference sample with heater temperature at 180°C.

(F) Measured temperature rising results for glass, metamaterial, and gold samples for different constant input heating power under atmospheric (ATM) and vacuum (VAC) conditions with a legend showing materials, conditions, and input power (voltage). Solid and dashed lines are ATM and VAC data results, respectively. The linewidth indicates different input power; y-axis is logarithmic.

(G) Micro-FTIR 2D mapping results of the integration measured emission signal for wavelength intervals of 2.5–5 µm, 5–8 µm, and 8–13 µm respectively from the left, with visible micro-image underneath for references.

A comparison of the emissivity for the original shape, rounded shape, and experimental measurement results are shown in Figure S4. The fabricated round shape does not significantly alter the emissivity profile since these small, rounded features are extremely small in scale compared to the IR-range wavelength. The deviation between the simulation and experimental measurement results may have resulted from fabrication imperfections and material impurities.

The surface pattern topography details of the metamaterial structure were characterized by atomic force microscope (AFM), as shown in Figure 6B, confirming that the overall fabricated pillar patterns exhibited consistent geometry with a well-defined vertical profile on top. The spectrum emissivity of the metamaterial structure was measured by Fourier transform infrared (FTIR) spectrophotometry (Shimadzu Iraffinity-1s), and the spectral radiant flux of the metamaterial structure at 180 °C is shown in Figure 6C with blackbody spectrum radiance at the same temperature for comparison. The spectral radiance indicates distinct emission enhancement within the 5–8 µm wavelength range and remarkably suppressing emission in the transparent atmospheric windows of 3–5 and 8–13 µm. This thermal emission suppression in the transparent atmospheric windows effectively minimizes the thermal signature of an object, making it less detectable by IR detectors. The enhanced emission in the 5–8 µm range is crucial for thermal management by promoting heat dissipation through thermal radiation.

IR thermal imaging of the metamaterial, alongside a glass substrate and a copper reference sample, was performed using a mid-wave IR (1.5–5.1 µm) thermal camera (FLIR X6530sc) and a long-wave IR (7.5–13 µm) thermal camera (FLIR SC620) with the heater at a stable temperature of 180 °C (Figures 6D and 6E). The thermal signature of the glass substrate closely matches the temperature of the heater because of its near-blackbody emissivity in both mid- and long-wave IR ranges. The copper reference sample showed a very low IR temperature for both mid- and long-wave IR detectors because of its low thermal emissivity in the IR range. Notably, the metamaterial structure exhibits a substantial IR temperature reduction compared to its actual temperature, showing nearly a 100 °C reduction for IR detectors. This significant reduction is attributed to the structure's engineered emission profile, which enhances emission within the 5–8 µm range while minimizing emission outside this band. Although the temperature difference between the metamaterial and copper references is noticeable because emitted radiation power is proportional to the fourth power of temperature, amplifying the IR temperature difference, the average emissivity of the metamaterial in the 3–5 and 8–13 µm ranges is about 0.091 and 0.103, respectively, slightly higher than that of copper (0.065).

To validate the radiative heat dissipation capability of the designed 5–8 µm channel, the metamaterial sample, along with glass and gold reference samples, was heated with different constant input power under atmospheric or high vacuum ($1 \times 10^4$ hPa) conditions to compare their temperature rise rates. The glass and gold reference samples were fabricated by depositing $SiO_2$ and Au thin films on silicon substrates of the same size as the metamaterial sample. As shown in Figure 6F, the metamaterial sample consistently exhibits a lower temperature than the gold reference sample, whereas the glass sample maintains the lowest temperature under the same conditions because of its highest radiative emission property for heat dissipation in the IR range. The temperature difference between the samples increases as the overall temperature rises. The rate of temperature increase accelerates with increasing input power, and the temperature differences between the gold, metamaterial, and glass samples become more pronounced because of the exponential increase in radiative heat flux with temperature. For the same power input, the temperature difference between the gold, metamaterial, and glass samples in vacuum conditions was significantly higher than in atmospheric conditions because thermal convection and conduction were reduced, making radiative heat transfer the primary mechanism for heat dissipation. Overall, the high emissivity of the metamaterial in the 5–8 µm range demonstrates its radiative heat management capability, particularly when compared to the gold reference, which emits negligible thermal radiation.

Figures 6G presents the 2D mapping results of the integrated emission signals measured by FTIR microspectroscopy (JASCO Co. IRT-5200/ FTIR-6600) for the wavelength intervals: 2.5–5 µm, 5–8 µm, and 8–13 µm respectively. The sample size is 200 by 100 µm$^2$ and the scanning step size is set as 10 µm to include sufficient metamaterial patterns for collective resonance effects. The "UT" logo micro-visible image is overlaid under each corresponding mapping result for reference. A strong integrated emission signal is clearly observed in the 5–8 µm range, whereas the signals in the 2.5–5 µm and 8–13 µm ranges are negligible. Noted that minor integrated emission signals at the edges were because of the residue of photoresist polymer. These micro-FTIR mapping results clearly demonstrate the emission enhancement within the 5–8 µm wavelength range, alongside suppression in the 2.5–5 µm and 8–13 µm ranges. Overall, the selectively tailored emissivity of the designed metamaterial demonstrates its potential for highly efficient thermal camouflage applications.

**Physical Mechanism Analysis**

This high emission property is enhanced by the excitation of surface plasmons and the coupling of magnetic polaritons resonance [63–68]. To understand the enhanced emission, three distinct peak wavelengths at 5.406, 6.488, and 7.241 μm, as indicated by the simulated spectrum radiance flux in Figure 6C, were investigated by electromagnetic field profiling for different cross-section areas. Because of the intrinsic high plasmon frequency and dense free electrons of the top gold layer, the structure can effectively block or reflect low energy IR light within the 3–15 μm range. The simulation is under *p*-polarized wave illumination for normal angle incidence since surface plasmons can only be excited by *p*-polarized waves.

For the emission peak $P_1$ at 5.406 μm, as seen from the electric field distribution for the *XY* plane at $Z = 0.01$ μm in Figure 7A, the surface plasmon resonance can be clearly observed by strongly enhanced and confined electric field distribution in the XY plane at the surface of the meta-structure. The electric field exhibits a "top–bottom" coupling pattern between adjacent patterns. The surface plasmon resonance peak wavelength is consistent with the periodic length. For the emission peak $P_2$ at 6.488 μm, as seen from both magnetic field distributions in Figure 7B, there are five distinct magnetic resonance patterns shown in both the *XY* and *XZ* cross-section planes, indicating high-order magnetic resonance. The strongly confined electric field in Figure 7B also implies the presence of surface plasmon polaritons. The strong magnetic field confinement in the middle spacer may indicate the coupling between surface plasmons and magnetic polariton resonance.

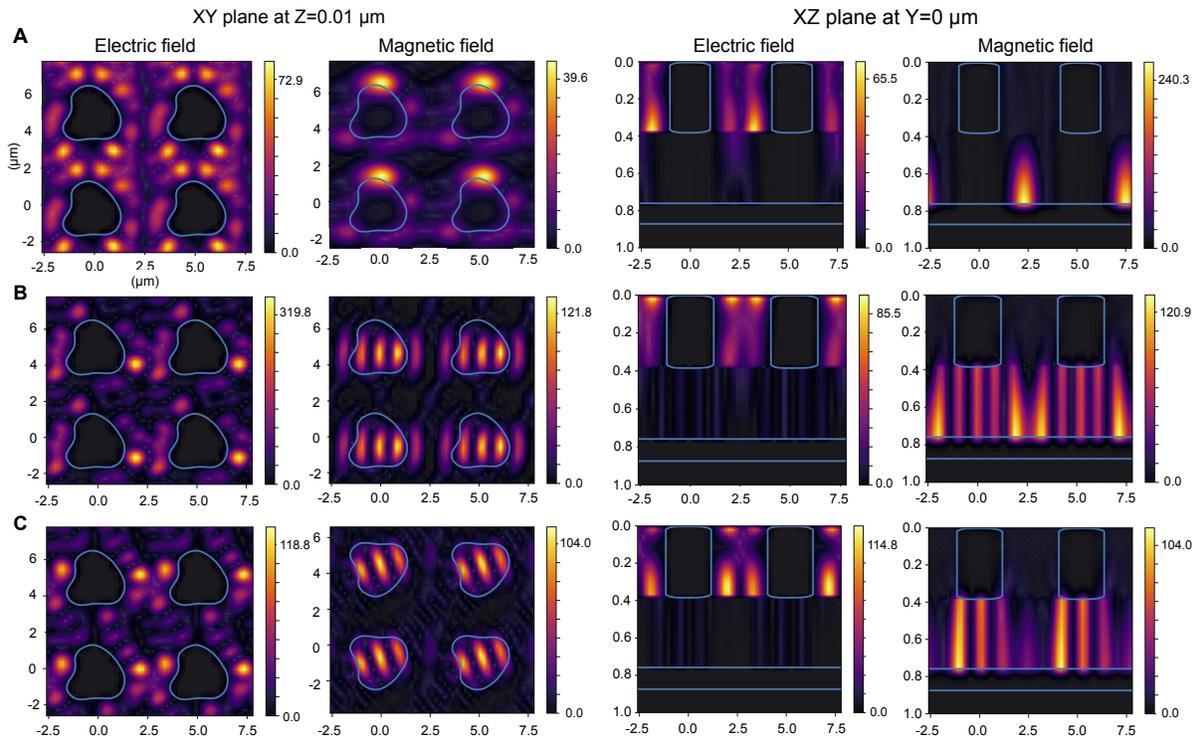

**Figure 7. Electromagnetic field distribution analysis.**
(A) 5.406 μm, (B) 6.488 μm, and (C) 7.241 μm. For each wavelength: The first two columns represent the electric field and magnetic field distribution for the XY plane at Z = 0.01 μm. The last two columns represent the electric field and magnetic field distribution for the XZ plane at Y = 0 μm.

The emission peak $P_3$ at 7.241 μm is also characterized by strongly enhanced and confined electric field dipoles in the *XY* plane at the meta-structure surface, as shown in Figure 7C. The electric field shows a "left–right" coupling pattern between adjacent patterns. The magnetic resonance signature can also be observed from the magnetic field in both the *XY* and *XZ* planes in Figure 7C, where three distinct magnetic resonance patterns indicate fundamental-order magnetic resonance. Similarly, the enhanced magnetic field between the top surface pattern and bottom metal reflector implies the coupling between surface plasmon and magnetic polariton resonances.

Compared to the standard circular shape [39–41,45,69,70], the free-form meta-surface pattern can excite asymmetrical surface plasmon resonance and coupling in different areas, such as top and bottom, and left and right coupling, which can support more modes to enhance and broaden the emission peak. Simultaneously, the coupling between surface plasmons and magnetic plasmon resonance can also promote emission enhancement.

## DISCUSSION

In this paper, we have proposed DiffMeta, a novel framework for the inverse design of metamaterials. Crucially, it produces a diverse variety of metamaterial structures and identify sensitive design parameters, beneficial for experimental fabrication. The fabricated free-form metamaterial designed by DiffMeta for thermal camouflage shows enhanced emissivity in the 5–8 μm range and high suppression of thermal emission signatures in the atmospheric windows at 3–5 and 8–14 μm, closely matching the target optical response. These findings suggest that DiffMeta can effectively address the complex and nonlinear relationships between metamaterial structures and their optical behaviours. By incorporating spectral conditioning mechanisms into the conditional diffusion model, DiffMeta significantly enhanced accuracy and diversity compared to VAE and GAN. Despite these promising results, one issue with DiffMeta is that it requires a large amount of data, time, and computational resources for training compared to other generative models. This high demand stems from the complexity of the DiffMeta model and its reliance on extensive data to capture the intricate distributions of the metamaterial patterns and properties.

## METHODS

### Metamaterial Structure Design

The MIM tri-layer metamaterial periodic structure was applied in this study, as shown in Figure 1A. The top layer was a free-form Au layer to excite and sustain multiple electromagnetic resonances, enhancing the emissivity property. The middle layer was an amorphous Si insulator thin film since silicon is transparent with a high refractive index in the wavelength range of interest. The bottom layer is a uniform Au thin film, used to block light transmission, enhance broadband IR light reflection, and support the magnetic resonance induced by the MIM structure. Fabrication constraints are imposed on several parameters: pitch size ($\phi_1$) from 3 to 8 μm, height of the top shape pattern ($\phi_2$) from 0 to 0.8 μm, middle dielectric spacer height ($\phi_3$) from 0 to 1 μm, and bottom reflector height ($\phi_4$) from 0 to 0.2 μm. The pattern shapes ranged from regular polygons to highly irregular forms, with the number of vertices limited to between 5 and 20. To accommodate fabrication limitations, the sharp features of the pattern were smoothed (see Supplemental Information Section S1.1 for design details).

### Spectrum Simulation

Optical emissivity is simulated using rigorous coupled wave analysis (RCWA) to obtain the reflection (*R*) and transmission (*T*) properties of the structures within the wavelength range of 3 to 15 μm. The absorption is calculated as (1-R-T) based on energy conservation, and emissivity is derived from the absorption property according to Kirchhoff's law for reciprocal materials (see Supplementary Section S1.2 for details).

### Sample Fabrication

The fabrication of the metamaterial structure is shown in Figure S2. Initially, Au (0.128 μm) and Si (0.381 μm) thin films were sequentially deposited onto a Si substrate using a radiofrequency (RF) sputtering

(CFS-4EP-LL). The sputtering deposition parameters are as follows: 200 W sputtering power, an Ar flow rate of 23 sccm/s, and sputtering pressure of $1 \times 10^{-3}$ Torr. Following the deposition, the substrate was spin-coated with polydimethylglutarimide (PMGI)-SF3 at 3000 rpm for 30 s. A pre-baking step was conducted at 180 °C for 180 s to cure the layers. Subsequently, a photoresist (JSR 7790G) was spin-coated at 3000 rpm for 30 s, resulting in a film thickness of approximately 1.1 μm. This layer was pre-baked at 110 °C for 90 s. The photomask design was performed using KLayout, and pattern transfer was accomplished using a HEIDELBERG DWL66+ laser lithography system using the high-resolution (HR) writing head.

After the laser writing, the photomask was developed for 45 s using an NMD-W developer, followed by three rinses with deionized water. The residual photoresist was removed using ozone plasma treatment for 15 s. Subsequently, a gold layer (0.389 μm) was deposited using the RF sputtering machine. The final step involved lifting off the photoresist using *N*-methyl pyrrolidinone (PG remover) at 80 °C and rinsing with deionised water several times to clean the substrate.

## RESOURCE AVAILABILITY

### Lead contact

Further information and requests for resources and reagents should be directed to and will be fulfilled by the lead contact, Junichiro Shiomi (shiomi@photon.t.u-tokyo.ac.jp).

### Data and code availability

The data generated in this study and the training data have been publicly deposited to Zenodo under (https://zenodo.org/records/12797962). The source code for DiffMeta, GANMeta, and VAEMeta used in this work is available at GitHub (https://github.com/tsudalab/DiffMeta).

## ACKNOWLEDGMENTS

This work was supported by the CREST Grant No. JPMJCR21O2 from the Japan Science and Technology Agency. R.H. acknowledges financial support from the JSPS Bilateral Joint Project (120227404) and National Natural Science Foundation of China (92463311, 52422603). The experimental nanofabrication was conducted at the Takeda Clean Room in the University of Tokyo, supported by "Advanced Research Infrastructure for Materials and Nanotechnology in Japan (ARIM)."

## AUTHOR CONTRIBUTIONS

Conceptualization, J.L., J.G., K.T., and J.S.; methodology, J.L., J.G., and K.T.; investigation, J.G.; writing—original draft, J.L., J.G., K.T., and J.S.; writing—review & editing, J.L., J.G., Z.M., D.D., J.S., T.X., Y.L., J.H., R.H., Y.Lee, K.T., and J.S.; funding acquisition, K.T. and J.S.; software, J.L. (diffusion model code) and J.G. (RCWA code); visualization, J.L. and Z.M.; validation, J.L. and J.G.; formal analysis, J.L.; fabrication, J.G., J.S., and Y.L.; temperature measurements, T.X.; IR imaging, J.H. and R.H.; experimental methodology, Y.Lee; supervision, K.T. and J.S.

## DECLARATION OF INTERESTS

The authors declare that they have no competing interests.

## SUPPLEMENTAL INFORMATION

**Document S1. Supplemental methods and Figures S1–S4**

## PROGRESS AND POTENTIAL

Machine learning is reshaping how we design materials by enabling inverse design, starting from a desired property and designing a structure to match it. This study presents a generative AI approach for metamaterial inverse design using conditional diffusion models. The framework generates diverse metamaterial structures tailored to a target selective emission spectrum and identifies key geometric features and fabrication parameters, helping researchers better understand how structure affects performance and guiding the experimental realization of a free-form metamaterial. The fabricated structure exhibits nearly 80% thermal emission at 180°C, while suppressing emissivity in infrared atmospheric windows, making it suitable for thermal camouflage. In the future, such AI-driven design tools could accelerate the development of advanced materials across energy, sensing, and defense technologies.